\documentclass[11pt]{article}
\usepackage{amsfonts,amsmath,amssymb}
\usepackage{color}
\usepackage{epic}
\usepackage{graphics}
\usepackage{boxedminipage}
\usepackage{epsfig}
\usepackage{changebar}
\usepackage{lscape}

\numberwithin{equation}{section}

\newcommand{\invdots}{{\mathinner{\mkern1mu\raise1pt
   \vbox{\kern7pt\hbox{.}}\mkern2mu\raise4pt\hbox{.}
   \mkern2mu\raise7pt\hbox{.}\mkern1mu}}}

\newtheorem{definition}{Definition}[section]

\newtheorem{proposition}[definition]{Proposition}

\newtheorem{corollary}[definition]{Corollary}

\newcommand{\prf}{\underline{Proof:}\ }
\newcommand{\finprf}{\null \hfill {\rule{5pt}{5pt}}\\ \null}

\definecolor{dcyan}{rgb}{0,.8,.8}
\definecolor{ddcyan}{rgb}{0,.6,.6}
\definecolor{dgreen}{rgb}{0,.8,0}
\definecolor{lyellow}{cmyk}{0,0,.2,0}
\definecolor{rose}{rgb}{1,.6,.6}
\definecolor{violet}{rgb}{.9,0,.6}

\newcommand{\1}{\mbox{\hspace{.0em}1\hspace{-.2em}I}}

\def\CC{{\mathbb C}}
\def\NN{{\mathbb N}}

\def\RR{{\mathbb R}}

          \def\cH{{\cal H}}          
                    \def\cL{{\cal L}}
\def\cM{{\cal M}}                    
                    
          \def\cT{{\cal T}}

\newcommand{\lda}{\lambda}

\newcommand{\ie}{{\it i.e.}\ }

\def\tU{{\widetilde{U}}}
\def\tV{{\widetilde{V}}}

\newcommand{\be}{\begin{eqnarray}}
\newcommand{\ee}{\end{eqnarray}}
\newcommand{\bea}{\begin{eqnarray}}
\newcommand{\eea}{\end{eqnarray}}

\begin{document}

\newpage
 \setcounter{page}{0}

\markright{\today\dotfill A multisymplectic approach to defects in integrable classical field theory \dotfill }
\pagestyle{myheadings}

\vspace{8mm}

\begin{center}
   {\Huge  {\sffamily A multisymplectic approach to defects\\
    in integrable classical field theory
  }}\\[1cm]

\vspace{8mm}

{\large \textbf{V. Caudrelier$^a$, A. Kundu$^b$}}\\
\vspace{1cm}
$^a$ Department of Mathematics, City University London, Northampton Square, EC1V 0HB London

 $^b$ Theory Division, Saha Institute of Nuclear Physics, 1/AF Bidhannagar,
Kolkata 700064, India

\end{center}

 \begin{abstract} We introduce the concept of multisymplectic formalism,
familiar in covariant field theory, for the study of integrable defects in
$1+1$ classical field theory.  The main idea is the coexistence of two
Poisson brackets, one for each spacetime coordinate.  The Poisson bracket
corresponding to the time coordinate is the usual one describing the time
evolution of the system.  Taking the nonlinear Schr\"odinger (NLS) equation
as an example, we introduce the new bracket associated to the space
coordinate.  We show that, in the absence of any  defect, the two brackets yield
completely equivalent Hamiltonian descriptions of the model.  However, in the
presence of a defect described by a frozen B\"acklund transformation, the
advantage of using the new bracket becomes evident.  It allows us to
reinterpret the defect conditions as canonical transformations.  As a
consequence, we are also able to implement the method of the classical $r$
matrix and to prove Liouville integrability of the system with such a
defect.  The use of the new Poisson bracket completely bypasses all the
known problems associated with the presence of a defect in the discussion of
Liouville integrability. A by-product of the approach is the reinterpretation of the 
defect Lagrangian used in the Lagrangian description of integrable defects as the generating function 
of the canonical transformation representing the defect conditions.
  \end{abstract}



\pagestyle{plain}
\setcounter{footnote}{0}
\section{Introduction}
Real materials and systems as a rule exhibit defect structures. 
However, since the presence of defects usually spoils the regular behaviour
predicted by the study of the ideal materials, in theoretical 
   investigations one usually tries  to avoid the  problem of defects,
which  in  general  is difficult to handle.
 Nevertheless, since the presence of defects, even a single one, might play
a pivotal role in determining bulk properties of a system, the investigation
of defect problems has been undertaken quite intensively in recent years,
both from theoretical and experimental point of view
\cite{mcCoy80}-\cite{jungwirth14}.

With the advent of experimental realisations of systems, the question of
defects,
 that has been
thought to be of purely academic interest
 for decades \cite{nanarro67}-\cite{vollhardt90},  emerged  naturally in
realistic models in  cold atoms and optical setups, which could be  described
efficiently  by $1+1$
dimensional integrable systems. 
 The initial focus was  mainly in quantum field theories \cite{DMS}-\cite{CMR} and
was concentrated until quite 
recently \cite{BG}-\cite{W} on various quantum systems. However,  it has
been understood soon, that in certain cases the presence of defects may be considered in an exact
manner, preserving even the integrability of the models. The general framework that includes
most of  the previous studies 
 was  proposed comprehensively in \cite{CMRS}.

 The question of integrable defects in classical field
theories was considered almost ten years after the publication of the first
paper on the topic in integrable QFT. 
In a series of papers \cite{BCZ}-\cite{CZ} related to several key
models like the sine-Gordon model, the NLS equation, etc, 
a Lagrangian approach was proposed where a contribution from the defect is
required to compensate
 for the loss of conservation of the momentum due to the  presence of a
defect. It was argued
that this is enough to ensure the  integrability of a defect model.
  A crucial observation to support this was that the conditions on the
fields that one obtains in this way correspond to B\"acklund transformations
frozen at the location of the defect.  This approach triggered a strong
activity in the analysis of the defect in integrable classical field
theories.
 The observation on frozen B\"acklund transformations was fully exploited in
\cite{VC} in conjunction with the Lax pair formulation of the general AKNS
approach \cite{AKNS} to obtain a generating function of the entire set of
modified conserved quantities.  This also allowed to answer some questions
left open in the Lagrangian formulation like the formulation of the defect
conditions directly in terms of the fields of the theory for models like KdV. 
It also settled the question of integrability in the sense of the presence
of an infinite
 number of conserved quantities.  But soon, the question of Liouville
integrability became a main issue.  The sine-Gordon model was the first
model to receive attention \cite{Hab-kundu}, followed by a very nice
series of papers tackling the question systematically for several models
\cite{avanDaik13,anvanSG10,Doiku-NLS}.  The procedure in these
investigations is based on the a priori assumption that the defect matrix
satisfies appropriate Poisson bracket relations formulated in the context of
the classical $r$-matrix approach.  A careful regularization is needed in
this procedure which yields the so-called ``sewing conditions" between
the fields in the bulk and those contained in the defect matrix.  The
consistency of the approach must then be checked a posteriori.

However, there  still exist two points of view, that have not been reconciled
so far.  On the one hand, the defect matrix (or operator) may be
given as a B\"acklund matrix involving the values of the fields
at the defect  point \cite{VC} but the Hamiltonian picture in this setting has not been understood so far.
On the other hand, one may start a priori with a Hamiltonian structure given by an $r$-matrix and require that the defect operator be given through a specific
realization of the corresponding Poisson algebra \cite{anvanSG10}. But then, it is not known how to connect this 
approach with that of the B\"acklund matrix. The two approaches, although 
linked to the same ideas, could not be unified easily. 
In the
first picture, one would like to {\it deduce} the Poisson brackets of the
defect matrix from its interpretation as a B\"acklund matrix, but this is
hindered by the divergence of the Poisson brackets at coinciding space
points.  In the second picture, the form of the Poisson brackets for the
defect matrix is {\it postulated} a priori using extra local fields, but it
then becomes difficult to eliminate those extra fields. In spite of the significant
success of these investigations, this issue is still open and represents a
wide
gap in the understanding of Liouville integrability for theories with a defect.

The purpose of this paper is to bridge  this gap by  reconciling these two
points of view, and at the same time to get rid of the limitations of each
approach by proposing a significantly new idea. Our approach also allows for a natural reinterpretation of 
 the defect density in the Lagrangian picture. Therefore, it provides a unifying framework of the three different approaches used so far for 
 the questions of integrable defects.
 The idea is to introduce an
additional Poisson structure in the theory, in terms of which the defect
conditions appear as canonical transformations.  This was the missing
ingredient in reconciling the various pictures and we show that the method of
the classical $r$-matrix and Liouville integrability with respect to
the \textit{ new Poisson structure}
 follows trivially 
from this observation.  
Interestingly, this  new
Poisson structure, which we call {\it equal-space} Poisson bracket, has been
known for a long time in other areas under the generic name of \textit{multisymplectic formalism}. 
Although this research area has developed in a rather non systematic 
 way (see e.g. \cite{BH} for an attempt to give an account of the various
approaches) and into a heavy mathematical formalism, the commonly accepted
origin is the so-called De Donder-Weyl formalism \cite{DDW}.  The basic
observation is that the traditional canonical formalism (either classical or
quantum) is grossly unbalanced in the way it treats time as opposed to the
other coordinates.  The De Donder-Weyl formalism, also called covariant
field theory formalism, aims at treating all independent variables on the
same footing.  We keep this simple idea and implement it directly in the
context of our defect problem,  on the
example of the NLS equation, showing how it combines nicely with the method
of the classical $r$-matrix.

The organization of the paper is as follows. In the next Section, we
summarize the motivation for introducing a multisymplectic structure in
field theory and introduce the relevant Poisson brackets for our purpose. 
It is shown how the usual classical r-matrix approach fits in the new
approach and we emphasize how the new structure brings in a completely
equivalent description for a system without a  defect.  In Section
\ref{defField}, we recall briefly the approach of \cite{VC} and then go on
to show how the new Poisson structure can be used to discuss Liouville
integrability and the classical $r$-matrix approach for the NLS model  with a defect. 
The last section contains our conclusions and perspectives on
future directions.

\section{Multisymplectic structure of the NLS equation}

\subsection{Space and time Hamiltonian forms for NLS}\label{multi_NLS}

Here we present a systematic account of the multisymplectic structure of the
NLS equation by introducing two Poisson brackets on the
phase space of the model.  This formalises the idea of ``dual picture"
discussed in \cite{AK}.  The main observation behind the multisymplectic
approach to field theory is that the canonical quantization procedure puts
emphasis only on the time parameter and, as a consequence, considers only  a
partial Legendre transformation when defining canonical conjugate
coordinates.  The traditional approach goes as follows.  Given fields
$\phi_a$ depending on coordinates $(x,t)$\footnote{For simplicity here, we
only consider two coordinates as this is enough for our purposes in $1+1$
dimensional field theory.} one defines the conjugate momenta $\pi^a$ as \bea
\label{Legendre1} \pi^a=\frac{\partial \cL}{\partial (\partial_t \phi_a)}\,,
\eea $\cL$ being the Lagrangian density.  Then, one imposes
\textit{equal-time} canonical relations by defining the {\it space} Poisson brackets as
\bea \{\phi_a(x,t_0),\pi^b(y,t_0)\}_S=\delta_a^b\delta(x-y)\,, \eea at some
initial time $t_0$, with the other brackets being trivial.  The subcript $S$ indicates that  
the Poisson bracket  is of  equal-time \ie it does not depend
on time but only on the space variables.  However, the Legendre transformation
(\ref{Legendre1}) is in fact incomplete, since   one can define also  another
complimentary set of
conjugate momenta as \bea \label{Legendre2} \Pi^a=\frac{\partial
\cL}{\partial (\partial_x \phi_a)}\,.  \eea The second ``dual" Poisson  bracket is then
defined in complete analogy by \bea
\{\phi_a(x_0,t),\Pi^b(x_0,\tau)\}_T=\delta_a^b\delta(t-\tau)\,, \eea at some
fixed location $x_0$, with  the other brackets being zero.  These relations
may
be seen as {\it equal-space} canonical brackets.  The subscript $T$
indicates that this Poisson bracket does not involve space variables.  These two
brackets can be combined and form the basis of the formulation of covariant
Poisson brackets for field theories.

We now discuss this idea in detail for the NLS and show that the two
brackets provide an equivalent description of the model.  Moreover, this
setting puts the two components of the Lax pair on an equal footing and the
coexistence of the two brackets is totally compatible with the usual
 properties of NLS model,  like the existence of an infinite number of
conserved quantities.  The main result of this section is, that the method of
the classical $r$ matrix, developed only for the bracket $\{~,~\}_S$, and based
on the space part of the Lax pair, goes over entirely to the new picture
where one uses $\{~,~\}_T$ and the time part of the Lax pair.  

We  emphasize strongly,  that the multisymplectic formalism presented here
is different from the well-known bi-Hamiltonian theory of integrable systems
\cite{Magri}.  The bi-Hamiltonian theory is based on the existence of two
compatible { equal-time } brackets $\{~,~\}_{S1}$ and $\{~,~\}_{S2}$, each
of which allows for the description of the {\it time} evolution of the
model.  Our equal-space bracket $\{~,~\}_{T}$ on the other hand is linked to
the {\it space} evolution of the model.  It is based on a completely
different Legendre transformation that is not considered in the traditional
canonical approach.

 Consider the NLS equation 
 \bea iq_t+q_{xx}=2\epsilon|q|^2q~,~~\epsilon=\pm
1 \,,\label{NLS} 
\eea for the complex field $q(x,t)$.  One can view the real
and imaginary parts of $q$ as the fields of the theory or equivalently can  take
$q\equiv \phi_1$ and its complex conjugate $q^*\equiv \phi_2$ as two
independent fields.  A Lagrangian density for this equation is \bea
\cL=\frac{i}{2}(\phi_2\phi_{1t}-\phi_{2t}\phi_1)-\phi_{2x}\phi_{1x}-\epsilon(\phi_2\phi_1)^2\,. 
\eea From this, we get \bea
\pi^1=\frac{i}{2}\phi_2~~,~~\pi^2=-\frac{i}{2}\phi_1\,, \eea 

One then obtains the NLS equation consistently\footnote{There is a subtlety here related to the fact that the Lagrangian for NLS 
is linear in the ``velocities" $\phi_{jt}$. Accordingly, one has to use the Dirac bracket. In our case, the net result of the standard approach to 
constrained systems is the factor $\frac{1}{2}$ in front of the bracket.}
as \bea
\pi^j_t=\frac{1}{2}\{\pi^j,H_S\}_S~~,~~j=1,2\,, \label{space_H_eq} \eea
where $H_S=\int\,\cH_S\,dx$ and
\bea 
\cH_S=\pi^1\phi_{1t} + \pi^2\phi_{2t}-\cL=\phi_{1x}\phi_{2x}+\epsilon \phi_1^2\phi_2^2
\,, 
\eea
The usual presentation of the Hamiltonian formulation of NLS takes a slight shortcut and considers the following canonical Poisson brackets\footnote{The 
reader will note the absence of a factor $2$ to compensate for the direct use of $\{~,~\}_S$ instead of the Dirac brackets.}
\bea 
\label{space_bracket}
\{q(x,t_0),q^*(y,t_0)\}_S=-i\delta(x-y)~~,
~~\{q(x,t_0),q(y,t_0)\}_S=0~~,~~\{q^*(x,t_0),q^*(y,t_0)\}_S=0\,. 
\eea
together with the Hamiltonian density 
\be
\label{Ham_S}
\cH_S=|q_x|^2+\epsilon|q|^4\,.
\ee
The equations of motion then read
\be
q_t=\{q,H_S\}_S\,.
\ee
 Now, in view of the above discussion on the equal-space canonical brackets,
we introduce new canonical conjugate fields to $\phi_j$ by \bea
\label{time_conjugate} \Pi^1=-\phi_{2x}~~,~~\Pi^2=-\phi_{1x}\,, \eea which
leads us to define the following new {\it equal space} Poisson brackets for NLS 
\bea
\label{time_bracket} &&\{q(x_0,t),q_x^*(x_0,\tau)\}_T
=-\delta(t-\tau)~~,~~\{q(x_0,t),q^*(x_0,\tau)\}_T=0=\{q_x(x_0,t),q_x(x_0,\tau)\}_T\,\nonumber
\\ &&\{q_x(x_0,t),q^*(x_0,\tau)\}_T
=\delta(t-\tau)~~,~~\{q(x_0,t),q(x_0,\tau)\}_T=0=\{q_x(x_0,t),q(x_0,\tau)\}_T,
\eea One then obtains the NLS equation from \bea \label{time_Ham_eq}
\Pi^j_x=\{\Pi^j,H_T\}_T~~,~~j=1,2\,, \eea where $H_T=\int\,\cH_T\,dt$ and,
\bea \label{Ham_T}
\cH_T=\Pi^1\phi_{1x}+\Pi^2\phi_{2x}-\cL=-|q_x|^2-\frac{i}{2}(q^*q_{t}-q^*_{t}q)+\epsilon|q|^4\,. 
\eea The new Hamiltonian density $\cH_T$ is the analog of $\cH_S$ with
respect to the new Poisson brackets $\{~,~\}_T$.  Indeed, a direct
calculation shows that the NLS equation (\ref{NLS}) is obtained from the Hamiltonian
equation \eqref{time_Ham_eq} with $j=2$ as \bea
-q_{xx}&=&\int\{-q_x(t),(-|q_x(\tau)|^2-\frac{i}{2}(q^*(\tau)
q_{\tau}(\tau)-q^*_{\tau}(\tau)q(\tau))+\epsilon|q(\tau)|^4)\}\,d\tau\\
&=&iq_t-2\epsilon|q|^2q\,.
\eea

\subsection{Classical $r$-matrix approach for the two Poisson brackets}\label{two_app}

\subsubsection{The standard approach with $\{~,~\}_S$}\label{standard}

The NLS equation arises as the compatibility condition of the auxiliary problem
\bea
\label{LP1}
\Psi_x(x,t,\lda)&=&U(x,t,\lda)\Psi(x,t,\lda)\,,\\
\label{LP2}
\Psi_t(x,t,\lda)&=&V(x,t,\lda)\Psi(x,t,\lda)\,,
\eea
with Lax pair
 \bea 
\label{LPUV} U=\left(\begin{array}{cc}
-i\lda &q\\
\epsilon q^* & i\lda
\end{array}\right)
~,~~V=\left(\begin{array}{cc}
-2i\lda^2-i\epsilon |q|^2& 2\lda q+iq_x\\ 
\epsilon(2\lda q^*-iq^*_x) & 2i\lda^2+i\epsilon|q|^2
\end{array}\right)\,.  
\eea 
The compatibility condition
$\Psi_{xt}=\Psi_{tx}$  results in the so-called zero curvature representation 
\begin{eqnarray} 
\label{zero_curv}
U_t-V_x+[U,V]=0\,,  
\end{eqnarray}
which must hold identically  for arbitrary  spectral parameter  $\lda$.

Using $U$ and $\{~,~\}_S$, one can develop the standard classical $r$-matrix
approach \cite{Sklyanin,FT} to discuss the Liouville integrability of the
model.  This approach is based entirely on the $x$-part of the Lax pair, eq. 
\eqref{LP1}, and time is considered as a fixed parameter, say $t=0$, which
is eventually evolved.  Therefore, we drop it in this section.  The starting
point of the method is the following ultralocal Poisson bracket
relation that may be derived
using the PB structure \eqref{space_bracket}:
 \bea \label{ultraloc}
 \{U_1(x,\lda),U_2(y,\mu)\}_S=\delta(x-y)\left[r(\lda-\mu),U_1(x,\lda)+
 U_2(y,\mu)\right]\,, \eea where we have used the notation $U_1=U\otimes
 \1$, $U_2=\1\otimes U$ and $r(\lda)$ is the usual $sl_2$ classical
 $r$-matrix \bea \label{r_matrix} r(\lda)=\frac{-\epsilon\,P}{2\lambda}\,,
 \eea $P$ being the permutation operator on $\CC^2\otimes \CC^2$: $P
 u\otimes v
=v\otimes u$.  We introduce the monodromy matrix $M_S(x,\lda)$ as the
fundamental solution of \eqref{LP1} (at $t=0$) equal to the identity matrix
at $x=0$.  Then, one computes for $x>0$, (see e.g.  \cite{FT})
 \bea \label{ybeUc}
 \{M_{S1}(x,\lda)\otimes,M_{S2}(x,\mu)\}_S=\left[r(\lda-\mu),M_S(x,\lda)\otimes
 M_S(x,\mu)\right]\,.  \eea On a finite interval $[0,L]$ with periodic
 boundary conditions, this relation is enough to conclude about the
 Liouville integrability of the model in the following sense: the transfer
 function $\cT_S(\lda)=trM_S(L,\lda)$ commutes for different spectral parameters
 \be
 \{\cT_S(\lda), \cT_S(\mu)\}=0 \ , 
 \ee 
 due to (\ref{ybeUc}), and hence generates the infinite set of
 conserved quantities $I_n$, $n\in\NN$, which are in  involution with respect to
 $\{~,~\}_S$ \be \{I_n,I_m\}_S=0~~,~~n,m\in\NN\,.  \ee
 These quantities can be extracted as local functionals of the
 fields $q$ and $q^*$ algorithmically by studying the large (real) $\lambda$
 expansion of the transfer function.  Full details are given in Section I.4
 of \cite{FT} and the net results is that, as $|\lambda|\to\infty$,
  \bea
  \arccos \left(\frac{1}{2}\cT_S(\lda)\right)+
\lambda L=\epsilon\sum_{n=1}^{\infty}\frac{I_n}{\lambda^n}\,,
  \eea
  where the integrals of motion $I_n$ are given by
  \be
  I_n=\int_0^L\,q^*(x,t)\,w_n(x,t)\,dx
  \ee
  and determined recursively using
  \bea
w_1=q~~,~~  w_{n+1}=-i\frac{\partial w_n}{\partial x}+\epsilon\,q^*\sum_{k=1}^{n-1}w_k\,w_{n-k}\,.
  \eea In particular, using $I_3$ (and an integration by parts) one extracts
 the Hamiltonian $H_S$ precisely as 
 \bea
 H_S=\int_0^L(|q_x|^2+\epsilon\,|q|^4)\,dx=\int_0^L\cH_S\,dx \eea where
 $\cH_S$ is the density given in \eqref{Ham_S}.  For our purposes, it
 is more convenient to arrive at the same result directly from the Lax
 pair formulation.  Representing $\Psi$ in \eqref{LP1}, \eqref{LP2} as a
 column vector \begin{eqnarray}
\Psi=\left(\begin{array}{c} \Psi^1 \\ \Psi^2 \end{array}\right)\,,
\end{eqnarray} 
and denoting $\Gamma=\Psi^2 \ (\Psi^1)^{-1}$, we derive
\bea
(\ln \Psi^1 )_x&=&U^{11}+U^{12}\Gamma \,,\\
 (\ln \Psi^1)_t&=&V^{11}+V^{12}\Gamma\,,
 \eea 
 where $U^{ij}$, $V^{ij}$ are the entries of $U$ and $V$. Then, $(\ln \Psi^1 )_{xt}=(\ln \Psi^1 )_{tx}$ yields the conservation equation
\begin{eqnarray} 
\label{conservation}
\left(U^{12}\Gamma\right)_t&=&\left(V^{11}+V^{12}\Gamma\right)_x\,, 
\ee 
since $U^{11}$ is constant. This shows using the periodic boundary condition,  that $\displaystyle\int_{0}^L U^{12}\Gamma\,dx$ is a generating function
of the conserved quantities.	
One can also derive a Riccati equation for $\Gamma$ 
\be 
\label{Riccati_x} 
\Gamma_x=2i\lda\Gamma+\epsilon\,q^*-q\Gamma^2\,, 
\ee
by using \eqref{LP1}. 
Expanding $\Gamma$ at $\lda\to\infty $: 
 \be 
 \label{GammaL}\Gamma=\epsilon\sum_{n=1}^\infty\frac{\Gamma_n}{(2i\lda)^n}\,, 
 \ee 
 and inserting in the Riccati equation, one gets a recursion relation for $\Gamma_n$
\be
\label{recursion1}
\Gamma_1=\,q^*~~,~~\Gamma_{n+1}=\Gamma_{nx}+\epsilon\,q\sum_{k=1}^{n-1}\Gamma_k\Gamma_{n-k}~~,~~n\ge1\,  
\ee 
 which allows to determine the successive integrals of motion $J_n$ as local functionals of the field and its space derivatives
 \be 
 \label{integrals1}
 J_n=\int_{0}^L q\Gamma_n\,dx~~,~~n\ge 1\,.
 \ee
 Note that $I_n^*=(i)^{n-1}J_n$ and one usually uses the combination
$\frac{1}{2}(I_n+I_n^*)$ to get real-valued conserved quantities. The connection just discussed between 
the $I_n$ and $J_n$ shows that the conserved quantities $J_n$ derived directly from the Lax pair presentation are in 
involution with respect to the Poisson structure $\{~,~\}_S$ introduced to describe NLS as a Hamiltonian system. 
\subsubsection{Classical $r$-matrix approach for the new bracket $\{~,~\}_T$}\label{class_new}
In this section, we show that one can formulate   a treatment  for
 the $t$-part of the auxiliary problem \eqref{LP2}, that goes completely
parallel to the usual classical $r$-matrix approach discuss above.  One has
to use the new Poisson bracket $\{~,~\}_T$ and the starting point is now an ultralocal relation involving the time Lax matrix $V$.  The
variable $x$ is a fixed parameter, say $x=x_0$, which could evolve
eventually. We have the following
\begin{proposition} Let the Poisson bracket $\{~,~\}_T$ be given by
\eqref{time_bracket} and $V$ be given by \eqref{LPUV}.  Then, \bea
\{V_1(t,\lda),V_2(\tau,\mu)\}_T=-\delta(t-\tau)
\left[r(\lda-\mu),V_1(t,\lda)+ V_2(\tau,\mu)\right] \,, \eea with the same
classical $r$-matrix as in \eqref{r_matrix}.  \end{proposition}
 \prf The
proof is essentially the same as for $U$ and follows from direct
computation.  We only give the main steps to illustrate the differences with
the usual computation.  Denote $U=-i\lambda \sigma_3+W, \ W=q \sigma^++
\epsilon q^* \sigma^-$ and then note that
$$V=-2i\lambda^2\sigma_3+2\lambda W-i\sigma_3W_x-iW^2\sigma_3.$$  In view of
\eqref{time_bracket}, the Poisson brackets $\{V_1(t,\lambda)\otimes
V_2(\tau,\mu)\}$ only involves four terms \begin{eqnarray*}
\{V_1(t,\lambda)\otimes
V_2(\tau,\mu)\}&=&-2i\lambda\{W_1(t),W_{x2}(\tau)\}(\1\otimes
\sigma_3)-2i\mu\{ W_{x1}(t) ,W_2(\tau)\}(\sigma_3\otimes \1)\nonumber\\
&&-\{W_1^2(t),W_{x2}(\tau)\}(\sigma_3\otimes \sigma_3)-\{ W_{x1}(t)
,W^2_2(\tau)\}(\sigma_3\otimes \sigma_3)\,.  \end{eqnarray*}
Performing  some algebra
yields \begin{equation*} \{V_1(t,\lambda)\otimes
V_2(\tau,\mu)\}=\epsilon\delta(t-\tau)\left[2i(\lambda+\mu)(\sigma_+\otimes
\sigma_- -\sigma_-\otimes \sigma_+) +(\1\otimes W-W\otimes
\1)(\sigma_3\otimes \sigma_3)\right]\,.  \end{equation*} On the other hand,
\begin{equation*}
[P,V_1(t,\lda)+V_2(t,\mu)]=2(\lambda-\mu)\left[i(\lambda+\mu)(\sigma_3\otimes
\1-\1\otimes \sigma_3)+(\1\otimes W-W\otimes \1)\right]P\,.  \end{equation*}
Noting that $(\sigma_3\otimes \1-\1\otimes \sigma_3)P=2(\sigma_+\otimes
\sigma_- -\sigma_-\otimes \sigma_+)$ and $(\1\otimes W-W\otimes
\1)P=(\1\otimes W-W\otimes \1)(\sigma_3\otimes \sigma_3)$ and using the
expression (\ref{r_matrix}) for the $r $-matrix , we get the result (see
also \cite{AK}).  \finprf
As a direct consequence, we obtain 
\begin{corollary}
\label{Coro_Mt}
 Let $M_{T}(t,\lda)$ be the fundamental solution of \eqref{LP2} (at $x=x_0$)
satisfying $M_T(0,\lda)=\1$, then for $t>0$, \bea \label{PBT}
\{M_{T1}(t,\lda),M_{T2}(t,\mu)\}_T=-\left[r(\lda-\mu),M_T(t,\lda)\otimes
M_T(t,\mu)\right]\,.  \eea \end{corollary} If we work on a finite time
interval $[0,\tau]$ with periodic conditions in time $q(x,0)=q(x,\tau)$, we
deduce that the transfer function $\cT_T(\lda)=trM_T(\tau,\lda), $ Poisson
commutes for different values of the spectral parameter.  We can therefore
talk about Liouville integrability of NLS in the same sense as before but
viewed with respect to $\{~,~\}_T$.  The transfer function $\cT_T(\lda)$
generates the conserved quantities (in space now) which are in involution
with respect to $\{~,~\}_T$.  To extract these conserved quantities, we
follow the same reasoning as in the previous section and use the
conservation equation \eqref{conservation}.  But this time, we interpret it differently, that is, 
as showing that
$\displaystyle\int_{0}^\tau\left(V^{11}+V^{12}\Gamma\right)\,dt$ is a
generating function for the conserved quantities in space.  Combined with
the following time-Riccati equation for $\Gamma$ \be \label{Riccati_t}
\Gamma_t=V^{21}+(V^{22}-V^{11})\Gamma-V^{12}\Gamma^2\,, \ee we obtain a
complete analog of the above algorithm for computing recursively the
conserved quantities.  Inserting the expansion \be
\Gamma=\epsilon\sum_{n=1}^\infty\frac{\gamma_n}{(2i\lda)^n}\,, \ee we obtain
\bea &&\gamma_1=- q^*~,~~\gamma_2=- q^*_x~,~ ~\gamma_3=-i q^*_t~- \epsilon
{q^*}^2 q, \\
&&~\gamma_{n+2}=i\gamma_{nt}+2\epsilon|q|^2\gamma_n+\epsilon
q\sum_{k=1}^{n}\gamma_k \gamma_{n+1-k}-\epsilon
q_x\sum_{k=1}^{n-1}\gamma_k\gamma_{n-k}~~,~~n\ge 1\,.  \eea Writing
$\displaystyle V^{11}+V^{12}\Gamma
=-2i\lda^2+\epsilon\sum_{n=1}^\infty\frac{{\cal K}_n}{(2i\lda)^n}$, the
corresponding integrals are \be K_n=\int_0^\tau {\cal K}_n\,dt=\int_0^\tau
i\left(q_x\gamma_n-q\gamma_{n+1}\right)\,dt\,.  \ee They are in involution
for the new Poisson bracket \be \{K_n,K_m\}_T=0~~,~~n,m\in\NN\,.  \ee In
particular, we find that \be \cH_T=\frac{i}{2}({\cal K}_2^*-{\cal K}_2)\,,
\ee
 where $\cH_T$ is the Hamiltonian density given in \eqref{Ham_T}. Hence, we
 recover the Hamiltonian $H_T$ precisely as \bea
 H_T=\int_0^\tau\cH_T\,dt=\int_0^\tau(-q^*_xq_x
 -\frac{i}{2}(q^*q_t-qq^*_t)+\epsilon(q^* q)^2)\,dt\,.  \eea
This concludes our presentation of the multisymplectic approach to NLS.
 \smallskip
\\
\noindent
\textit{Remark:} The choice of periodic boundary conditions in time is
 solely motivated by the need to keep the discussion as concise as possible
 at the technical level.  Of course, by analogy with the usual ``space" case,
 one could consider ``open" boundary conditions at $t=0$ and $t=\tau$.  In
 that case, the analog of Sklyanin's theory for systems on an interval
 should be implemented.  This can obviously be done here since the
 fundamental algebraic structure is the same.  Another possibility would be
 to consider vanishing conditions at infinity.  Again, there is no deep
 obstacle to this.  The usual class of solutions obtained from initial
 conditions satisfying $\displaystyle\lim_{|x|\to\infty} q(x,0)=0$ contains
 for instance the well-known $N$-soliton solutions.  These solutions are in
 fact well-defined for all $t\in\RR$ and satisfy
 $\displaystyle\lim_{|t|\to\infty}q(x_0,t)=0$ for arbitrary but fixed $x_0$. 
 Therefore, it would make sense to consider the \textit{time} problem on the
 line with vanishing boundary conditions at infinity.
\subsection{Canonical transformations and B\"acklund transformations}\label{canonical_Backlund}
We recall some elementary facts about canonical transformations in the
Hamiltonian formalism.  The main message from Sections \ref{multi_NLS} and
\ref{two_app} is that there is a complete duality in the structures and the
Hamiltonian formalism for NLS whether one uses the (usual) space point of
view or the (new) time point of view.  Therefore, all we have to do to
discuss canonical transformations for NLS simultaneously for $\{~,~\}_S$ and
$\{~,~\}_T$ is to use a generic Poisson bracket $\{~,~\}_Z$ and canonical
fields $Q_j$, $P^j$ of the independent variables $u,v$.  The results will
apply to each situation simply by performing the following identifications:
\begin{itemize}
\item Traditional approach:
\bea
\label{tradi_app}
~\{~,~\}_Z=\{~,~\}_S~~,~~Q_j=\phi_j~~,~~P^j=\pi^j~~,~~u=x~,~v=t~\text{(with $t$ fixed)}\,,
\eea
\item New approach: \bea \label{new_app}
\{~,~\}_Z=\{~,~\}_T~~,~~Q_j=\phi_j~~,~~P^j=\Pi^j~~,~~u=t~,~v=x~\text{(with
$x$ fixed)}\,.  \eea \end{itemize} Given our purposes below in connection
with canonical properties of B\"acklund transformations, we follow the
method of \cite{Wadati,Kodama} generalizing it to our new approach and consider canonical transformations
$\{Q_j,P_j\}\to \{\widetilde{Q}_j,\widetilde{P}_j\}$ that preserve the form
of the local conserved densities ${\cal I}_n$ of the theory, in the sense
that there should exists functionals ${\cal F}_n$ such that \be
\label{infinite_canonical} \widetilde{{\cal I}}_n={\cal I}_n+\partial_u
{\cal F}_n\,.  \ee For the conserved quantities, this yields \be
\label{constant_difference} \widetilde{ I}_n=I_n+E_n\,, \ee where  $E_n$
are constants obtained by integrating  $\partial_u {\cal F}_n$ on
the relevant interval $U$ for $u$.  Of course, in the traditional approach
where $u=x$ and $U$ is either $\RR$ (with vanishing conditions
for the fields and their derivatives  at infinity) or the interval $[0,L]$ (with periodic
boundary conditions), these constants are zero.  This is a natural
generalization to the case of integrable systems of the usual notion of
 canonical transformations that are required to preserve the form of the
Hamiltonian \be \widetilde{H}=H+E\,, \ee where the constant $E$ comes from
the fact that one considers so-called restricted canonical transformations. 
More precisely, one requires that the one-forms representing the system in
old and new variables differ only by a exact form \be
\int_U\,du(\widetilde{P}^j\,d\widetilde{Q}_j)-\widetilde{H}\,dv=\int_U\,du(P^j\,dQ_j)-H\,dv+dF\,,
\ee where $F$ is the so-called generating functional and is taken to be  as
 \be \label{split_cano}
F[Q_j,P^j,\widetilde{Q}_j,\widetilde{P}^j,v]=S[Q_j,P^j,\widetilde{Q}_j,\widetilde{P}^j]-Ev\,. 
\ee Assuming that the new variables do not depend explicitely on $v$, we get
the well-known transformation formulas \be \label{canonical_transfo}
P^j=\frac{\delta F}{\delta Q_j}~~,~~\widetilde{P}^j=-\frac{\delta F}{\delta
\widetilde{Q}_j}\,, \ee where we have assumed that $Q_j$ and
$\widetilde{Q}_j$ were  functionally independent variables (corresponding
to the so-called type $1$ generating functional).
\smallskip
\\
In the traditional approach \eqref{tradi_app}, the above discussion, in
particular eq.  \eqref{infinite_canonical}, was used in \cite{Kodama,Wadati}
to show that B\"acklund transformations naturally arise as canonical
transformations of the restricted type considered here.  For clarity and
self-containedness, we rewrite here the main line of arguments but expressed  directly
in the Lax pair formalism.  Since we want to preserve the form of the
Hamiltonian (and hence the equation of motion) as well as that of  all
 conserved quantities, we look for a transformation that preserves the
zero curvature representation of the equation of motion.  Looking at the
$x$-part only of the auxiliary problem   \eqref{LP1}, for fixed $t$,  we
introduce a matrix $L$ such that
$\widetilde\Psi(x,\lda)=L(x,\lda)\Psi(x,\lda)$, and satisfying \be
\label{eq_L} \partial_x L=\widetilde{U}L-L\,U\,, \ee where $\widetilde{U}$
is of the same form as $U$ but with $\phi_j$ replaced with
$\widetilde{\phi_j}$.  As explained in Section \ref{standard}, the infinite
set of conserved quantities is generated by \be
J(\lda)=\int_0^LU^{12}\Gamma(\lda)\,dx\,, \ee and similarly for the
conserved quantities in the new variables \be
\widetilde{J}(\lda)=\int_0^L\widetilde{U}^{12}\widetilde{\Gamma}(\lda)\,dx\,,
\ee with obvious notations.  Therefore, \eqref{infinite_canonical} will be
fulfilled if we can write \be
\widetilde{U}^{12}\widetilde{\Gamma}(\lda)=U^{12}\Gamma(\lda)+\partial_x
{\cal F}(\lda)\,, \ee for some functional ${\cal F}$.  Now using the
definition for $L$, \eqref{eq_L} and the Riccati equation \eqref{Riccati_x},
we find \be \label{exact_diff}
\widetilde{U}^{12}\widetilde{\Gamma}(\lda)=U^{12}\Gamma(\lda)+\partial_x
\ln(L^{11}+L^{12}\Gamma(\lda))\,.  \ee hence ${\cal
F}=\ln(L^{11}+L^{12}\Gamma)$, and the B\"acklund transformation associated
to $L$ is canonical.  A consequence of this result is that the Lax matrix
$\widetilde{U}(x,\lda)$ also satisfies the ultralocal relations
\eqref{ultraloc}.  Indeed, a standard argument shows that the new variables
$\widetilde{\phi}_j$ and $\widetilde{\pi}_j$ satisfy the same canonical
Poisson brackets as $\phi_j$ and $\pi_j$, \ie \eqref{space_bracket} in the
present case.  The method of the classical $r$-matrix can then be used
entirely for the new variables $\widetilde{\phi}_j$ and
$\widetilde{\pi}_j$.
\\
The adaptation of this reasoning to the new approach \eqref{new_app}, done
in the next section, is the key in reinterpreting defect conditions arising
from frozen B\"acklund transformations at a fixed location $x=x_0$ as
canonical transformations of the system.  As a consequence, we will be able
to conclude on the Liouville integrability of NLS with such defect
conditions.

\section{NLS with a defect: Liouville integrability}
\label{defField}
\subsection{Defects as frozen B\"acklund transformations}
 Viewing a defect in space as an internal boundary condition on the fields and
their time and space derivatives at a given point, the fruitful idea of
{\it frozen} B\"acklund transformations,
 originally noticed in \cite{BCZ}, is a convenient way of introducing
 integrable defects in classical field theories described by a Lax pair. 
 The systematic procedure for a large class of integrable classical field
 theories was described and implemented in \cite{VC}, where a generating
 function for the defect contributions to the conserved quantities was
 explicitely constructed.  This allows  to speak of the integrability of such
 defect conditions in the sense of the existence of an infinite number of
 conserved quantities.
\smallskip
\\
The main steps go as follows. Consider another copy of the
auxiliary problem for $\widetilde{\Psi}$ with Lax pair $\widetilde{U}$, $\widetilde{V}$ defined as
in (\ref{LP1},\ref{LP2}) with the new field $\widetilde{q}$ replacing $q$.  
  We fix a point $x_0\in[0,L]$ and use the auxiliary problem (\ref{LP1},\ref{LP2}) to describe the system $x>x_0,$ while the one with $\tU$
and $\tV$ describe the system for $x<x_0$.  At $x=x_0$, the two systems are connected via the condition
\begin{eqnarray} 
\label{def_L}
\widetilde{\Psi}(x_0,t,\lda)&=&L_0(t,\lda)\,\Psi(x_0,t,\lda)\,.
\end{eqnarray}
In turn, this yields the defect conditions in the form
\begin{eqnarray}
\label{eq_diff_L2}
L_{0t}(t,\lda)&=&\widetilde{V}(x_0,t,\lda)\,L_0(t,\lda)-L_0(t,\lda)\,V(x_0,t,\lda)\,,
\end{eqnarray} 
where we have denoted $\displaystyle V(x_0,t,\lda)=\lim_{x\to x_0}V(x,t,\lda)$ and similarly for $\widetilde{V}$.
The matrix $L$ is called the \textit{defect matrix}. 
With this construction, one can identify the
generating function of the defect contribution to the conserved quantities as follows \cite{VC}.  
 \begin{proposition}
 \label{prop_conserved}
The generating function for the integrals of motion
reads
 \begin{eqnarray} 
 \label{generating}
I(\lda)&=&I_{bulk}^{left}(\lda)+I_{bulk}^{right}(\lda)+I_{defect}(\lda)\,,
\ee 
where
\be 
I_{bulk}^{left}(\lda)&=&\int^{x_0}_{0}
\widetilde{U}^{12}\widetilde{\Gamma}dx\,,\ \
I_{bulk}^{right}(\lda)=\int_{x_0}^L U^{12}\Gamma dx\,,\\
I_{defect}(\lda)&=&\ln (L^{11}+L^{12}\Gamma)|_{x=x_0}\,, 
\label{Idefx} \end{eqnarray} and
$L^{ij}$'s are the entries of the defect matrix $L$. This means that 
\be
\partial_t\,I(\lda)=0\,.
\ee 
\end{proposition} 
The previous result gives the generating function of the infinite set of
modified conserved quantities (in time) and can be combined with
\eqref{recursion1} and \eqref{integrals1} to extract them order by order. 
From the point of view of PDEs, one can then speak of integrability, though   from
the point of view of Hamiltonian integrable systems, the question of
Liouville integrability is still not solved.  This is done in the next Section.
\subsection{Liouville integrability: defect conditions as canonical transformations}
The discussion of Section \ref{canonical_Backlund} and the short review in
the previous section about defect conditions arising from frozen B\"acklund
transformations make it plain that such defect conditions are nothing but
canonical transformations for the new bracket $\{~,~\}_T$.  Indeed, one can
repeat word for word the arguments of Section \ref{canonical_Backlund} but
using the new approach \eqref{new_app} instead of the usual one
\eqref{tradi_app}.  The key equations \eqref{eq_L} and \eqref{exact_diff}
are replaced respectively by \eqref{eq_diff_L2} and 
\be \label{rel_t}
\widetilde{V}^{11}+\widetilde{V}^{12}\,\widetilde{\Gamma}=V^{11}+V^{12}\,\Gamma+\partial_t
\ln(L_0^{11}+L_0^{12}\Gamma)\,.  
\ee Here, the time Riccati equation
\eqref{Riccati_t} should be used in establishing this last result. 
Comparing \eqref{rel_t} with the general discussion of Section
\ref{canonical_Backlund}, we see that, denoting  ${\cal E}_2$ as the
coefficient of $\lda^{-2}$ in the expansion of
$i(\ln(L_0^{11}+L_0^{12}\Gamma)^*- \ln(L_0^{11}+L_0^{12}\Gamma))$, we obtain
\be 
\label{ex_H_T}
\widetilde{H}_T=H_T+\left[{\cal E}_2\right]_0^\tau\,.  
\ee The
canonical transformation formulas \eqref{canonical_transfo} also allow us to
reinterpret the defect lagrangian density originally introduced in the
Lagrangian approach to integrable defects \cite{BCZ} as the density for the
generating functional of the canonical transformation.  This provides an
explicit check that the frozen B\"acklund defect conditions are indeed
canonical transformations with respect to our Poisson structure. For instance, from the defect density given in \cite{Z} (eq. (3.2)), in the focusing case $\epsilon=-1$, we
find that by choosing $S$ in \eqref{split_cano} as 
\be
S[\phi_j,\widetilde{\phi}_j,\Pi^j,\widetilde{\Pi}^j]= \int_0^\tau
\left(\frac{i\Omega}{2}\partial_t\ln\left(\frac{\widetilde{\phi}_1-\phi_1}
{\widetilde{\phi}_2-\phi_2}\right)+\frac{\Omega^3}{3}
+\Omega(\widetilde{\phi}_1
\widetilde{\phi}_2+\phi_1\phi_2-\alpha^2)-i\alpha(\widetilde{\phi}_2\phi_1-\widetilde{\phi}_1\phi_2)
\right)dt\,, \ee where \bea
\Omega&=&\pm\sqrt{\beta^2-(\widetilde{\phi}_1-\phi_1)(\widetilde{\phi}_2-\phi_2)}\,,
\eea then, after some algebra, eqs \eqref{canonical_transfo} yield the
following defect conditions at $x=x_0$ \be \begin{cases}
\widetilde{\phi}_{1x}-\phi_{1x}
=i\alpha(\widetilde{\phi}_1-\phi_1)+(\widetilde{\phi}_1+\phi_1)\Omega\,,\\
\widetilde{\phi}_{1t}-\phi_{1t}
=-\alpha(\widetilde{\phi}_{1x}-\phi_{1x})+i(\widetilde{
\phi}_{1x}+\phi_{1x})\Omega+i(\widetilde{\phi}_{1}-\phi_{1})
(\widetilde{\phi}_1\widetilde{\phi}_2
+\phi_1\phi_2)\,, \end{cases} \ee upon
recalling that $\widetilde{\Pi}_2=-\widetilde{\phi}_{1x}$,
$\Pi_2=-\phi_{1x}$.  These are precisely the defect conditions found in
\cite{VC}, $\alpha$ and $\beta$ being two arbitrary real numbers known to
parametrise the B\"acklund transformation for NLS.
\\ \\
At this stage, it is important to analyse what we have just obtained and our
claim of Liouville integrability of the  NLS model  with a  defect.  The point is that, for
$x\in[0,x_0)$, we describe the NLS model  in the bulk using the new Poisson bracket
$\{~,~\}_T$ and the associated transfer function ${\cal T}_T(\lda), $ which
ensures that the system is Liouville integrable, as discussed in detail in
Section \ref{class_new}.  At $x=x_0$, we simply reinterpret the defect
conditions as a passage from the old canonical variable $\phi_j$, $\Pi_j$
(see \eqref{time_conjugate} and \eqref{time_bracket}) to new canonical
variables $\widetilde{\phi}_j$, $\widetilde{\Pi}_j$.  From the point of view
of the new bracket, the defect conditions are simply a canonical change of
variables used to describe the system.  The bulk system for $x\in(x_0,L]$ is
then the result of the space evolution from $x=x_0$ to $x=L$ of the system
described in the new variables.  Therefore, we simply have to apply the
results of Section \ref{class_new} in the new canonical variables to
conclude about the Liouville integrability of the system.  In particular,
$\widetilde{\cal T}_T(\lda)$ generates the conserved quantities that are in
involution.  By construction, the new conserved quantities $\widetilde{K}_n$
only differ from the old ones $K_n$ by constants (which vanish under our
assumptions of periodicity in time).  For instance, as seen above at the
level of the Hamiltonian, the corresponding constant is $\left[{\cal
E}_2\right]_0^\tau$.  Therefore, at $x=x_0$, one is free to use either the
old or the new canonical variables to describe the system.  The difference
between the two canonical pictures is entirely encoded in the defect matrix
$L$.  This will be even more transparent in the next section when we
introduce the monodromy matrix of the system with defect.  At the location
of the defect, we will have two equivalent options to represent the
monodromy matrix (see  \eqref{def_mono_matrix}
 below).
\subsection{Liouville integrability: classical $r$-matrix approach with defect}
Let us first recall the problem that one encounters, when one wants to
generalise directly the $r$ matrix approach of (\ref{ybeUc}) based on $M_S$
and $\{~,~\}_S, $ to the case with an integrable defect of the type
described above.  From this point of view, the natural object to consider is the monodromy matrix
 $\cM_S(\lda)\equiv \widetilde{M}_S(0,x_0,\lda)\,L_0(0,\lda)\,
M_S(x_0,L,\lda)$, where we
have added an argument in $\widetilde{M}_S$ and $M_S$ to specify what part
of the interval $[0,L]$ they encode.  The problems then come in two
flavours.  First, there is a serious technical difficulty appearing when one
wants to compute the Poisson bracket $\{\cM_S(\lda),\cM_S(\mu)\}_S$ due to
the fact that, in this picture, $L_0$ is expressed in terms of the fields at
\textit{coinciding} points in space (the location $x_0$ of the defect). 
This can be overcome with great effort thanks to  a discretization procedure as
described in \cite{avanDaik13}. But then, there is a conceptual difficulty: one replaces
the fields of the bulk evaluated at the location of the defect by new local fields inside $L_0$. Then,
the desired form of the Poisson bracket involving $L_0$ with itself is
postulated ad hoc ``to make things work", hence imposing the Poisson brackets of the local degrees of
freedom living at the defect location. In other words, one simply assumes that $L_0$ satisfies the Poisson
algebra \eqref{ybeUc}.  Then, one checks a posteriori, that this is
consistent, which gives
rise to the ``sewing conditions" of \cite{avanDaik13} between the bulk fields and the defect fields.
\\
As we have argued above, the use of the new Poisson brackets to discuss the
Hamiltonian structure of NLS with a defect allows us to reinterpret the
defect conditions simply as a canonical transformation, whereby one decides
to change the variables used to describe the system at a specific point in
space and then lets the system evolve (in space) in the new canonical
variables.  We now show, that this new point of view allows us to solve the
above problems in a natural way.  Liouville integrability with a defect, already
established in the previous section, is then also manifested through the
classical $r$-matrix formalism.  In our setting, the natural object to
consider is the monodromy matrix $\cM_T(x,t,\lda)$ analogous to $M_T$ of
Section \ref{class_new}, but which takes into account the fact that we change
the variables at the location $x=x_0$.  It is given by \bea
\label{def_mono_matrix} \cM_T(x,t,\lda)=\begin{cases}
\widetilde{M}_T(t,\lda)~~,~~0\le x<x_0\,,\\
\widetilde{M}_T(t,\lda)=L_0(t,\lda)M_T(t,\lda)~~,~~x=x_0\,,\\
M_T(t,\lda)~~,~~x_0<x\le L\,.  \end{cases} \eea where $M_T$ is the matrix
considered in Corollary \ref{Coro_Mt} and $\widetilde{M}_T$ is the analogous
matrix but constructed from the new canonical variables.  So we obtain
immediately, for all $x\in[0,L]$, \bea \label{brackets_time_mono}
\{\cM_{T1}(x,t,\lda),\cM_{T2}(x,t,\mu)\}_T=-\left[r(\lda-\mu),\cM_{T1}(x,t,\lda)\,
\cM_{T2}(x,t,\mu)\right]\,.  \eea The transfer matrix of the system with
defect is now \be {\cal T}^d_T(\lda)=Tr \cM_T(x,\tau,\lda)\,, \ee and it
generates the conserved quantities which are in involution with respect to
$\{~,~\}_T$.
\section{Concluding remarks} 
Using the example of NLS, we have introduced the notion of multisymplectic
formalism in the context of integrable classical field theory.  This was
motivated by an unresolved issue in the area of integrable defects in
classical field theory.  We showed the complete equivalence between the
usual space canonical approach and new time canonical formulation for NLS
equation without a defect.  The equivalence goes over to the associated
classical r-matrix approaches.  The advantage of using the new Poisson
structure becomes apparent when one incorporates integrable defects in the
form of frozen B\"acklund transformations at a specific location.  Indeed,
with respect to the new structure, these can be reinterpreted as canonical
transformations and one can immediately conclude on the Liouville
integrability of the system with such a defect.  Once again, this goes over
to the classical r matrix formalism where the presence of the defect is
absorbed in the time monodromy matrix by changing the canonical variables
used to describe the system.  This clarifies the missing picture between the
B\"acklund approach to integrable defects and the (standard) classical r
matrix appraoch.
\\
Let us remark that we have three related interpretations for an integrable defect, each of which 
having its advantages, that are now unified in our picture. Historically, in the Lagrangian appraoch of \cite{BCZ}, the 
defect was introduced as a set of boundary conditions chosen so as to ensure that certain quantities are restored 
as conserved quantities once the defect contribution is taken into account. The prime example was momentum which is lost a priori 
due to the breaking of translational symmetry. Then, it was noticed by the same authors that the boundary conditions they found this way 
were (frozen) B\"acklund transformations. This observation was used extensively in \cite{VC} to discuss the generating functional of the 
entire hierarchy of conserved quantities which are known to be a dual facet to the symmetry content of a system. Finally, in the 
present paper, a third interpretation is presented whereby these boundary conditions appear as canonical transformations of a certain type 
with respect to the new Poisson bracket we introduced. This was possible because we put both space and time coordinates on the same footing and 
considered a Poisson bracket corresponding to space evolution. By changing the roles of space and time, the defect boundary conditions now correspond 
to a canonical transformation of the fields. From this point of view, the conserved quantity (and therefore the symmetry) content ``before" 
and ``after" the transformation (\ie on one side of the defect location and then on the other side) is the same. This can be seen for instance from eq \eqref{rel_t}
 and its consequence 
\eqref{ex_H_T} which show that the old and new conserved quantities only differ by constants which vanish under appropriate (time) boundary conditions. 
The symmetry content is not affected by the present B\"acklund type canonical transformations. In fact, this could be taken as an explanation of the 
integrable nature of such defect conditions.
 \\
This new theory will apply to other integrable classical field models
(for instance the sine-Gordon model) or even discrete integrable models
(like the Toda  chain).  We hope to return to these in the near future.
Our result also opens, in principle, the way  to the quantization of such
models with defects.  Indeed, when reformulated with the new Poisson
structure, it is clear that the canonical property of the defect conditions
is related to their origin in the form of B\"acklund transformations.  On
the other hand, the r matrix appearing in the time brackets is the same as
that appearing in the usual space brackets.  Therefore, it will be an
interesting problem to understand how the quantization of these brackets and
the quantization of B\"acklund transformations continue to intertwine so as
to produce a quantum integrable system with integrable defect conditions. 
It is all the more interesting as quantum B\"acklund transformations are
usually understood in connection with Baxter's $Q$ operator \cite{PG,SK}.

\end{document}